\documentstyle[preprint,aps,prl,psfig]{revtex}
\tighten
\begin{document}

\title{The $\beta$-relaxation dynamics of a simple liquid}
\author{Tobias Gleim and Walter Kob}
\address{Institute of Physics, Johannes Gutenberg-University,
Staudinger Weg 7, D-55099 Mainz, Germany}
\maketitle

\begin{abstract}
We present a detailed analysis of the $\beta$-relaxation dynamics of a
simple glass former, a Lennard-Jones system with a stochastic dynamics.
By testing the various predictions of mode-coupling theory, including
the recently proposed corrections to the asymptotic scaling laws, we
come to the conclusion that in this time regime the dynamics is
described very well by this theory.

\begin{center} January 29, 1999 \end{center}
\end{abstract}


\noindent
In the last few years it has been demonstrated that mode-coupling
theory (MCT)~\cite{mct} is able to describe many aspects of the
relaxation dynamics of supercooled liquids. In particular the theory is
able to explain on a qualitative level, and for certain systems even on
a quantitative one, phenomena like the non-Debye behavior of the
$\alpha$-relaxation process, the wave-vector dependence of the
Lamb-M\"ossbauer and Debye-Waller factors, and why quantities like the
viscosity or the $\alpha$-relaxation times show an anomalously strong
temperature dependence of their activation energy in the vicinity
of $T_c$, the so-called critical temperature in the theory.

Apart from the existence of $T_c$, the most important predictions of
MCT deal with the so-called $\beta$-relaxation process which is
proposed to exist in the supercooled regime on the time scale between
the microscopic relaxation at short times and the $\alpha$-relaxation
at long times. The $\beta$-regime is readily seen if a time correlation
function $\phi(t)$, such as the intermediate scattering function, is
plotted versus the logarithm of time. In the supercooled regime
$\phi(t)$ will show at intermediate times a plateau, and the relaxation
dynamics of the system on the time scale at which $\phi(t)$ is close to
this plateau is the $\beta$-regime.  The reason for the existence of
this plateau is that on this time scale the particles are trapped in
the cages formed by their surrounding neighbors. Hence the predictions
of the theory regarding the $\beta$-regime deal with the details of the
dynamics of the particles in these cages. Some of these predictions
have already been confirmed by various experiments on colloidal
suspensions and molecular liquids~\cite{glass_proceedings}.

In the past it has been shown that apart from experiments also computer
simulations are a very useful tool to probe the dynamics of supercooled
liquids~\cite{kob99}.  Because simulations allow the investigations of
observables which in real experiments are hard to measure, as e.g. the
dynamics at large wave-vectors or cross-correlation functions, they
permit to make more stringent tests of theoretical concepts and thus
are a valuable addition to experiments.  Results of such tests have,
e.g., been done for soft sphere systems~\cite{roux89}, Lennard-Jones
models~\cite{kob_lj}, water~\cite{sciortino}, and
polymers~\cite{bennemann98}. The result of these
tests was that the theory is indeed able to give a good description of
the relaxation dynamics of these systems. What these simulations have,
however, not been able to address so far are several important
predictions, discussed below, of MCT about the relaxation dynamics in
the $\beta$-regime. The main reason why these predictions have not been
tested was that they are supposed to be valid only very close to the
critical temperature $T_c$ of MCT, and that close to $T_c$ the
relaxation times of the system are usually so large that it is very
hard to equilibrate the system within the time span accessible to a
computer simulation (but easily reachable in a real experiment). If the
predictions of the theory are tested at slightly higher temperatures,
where the system can be equilibrated even in a computer simulation, the
strong interference of the microscopic dynamics of the system with the
$\beta$-relaxation process will spoil the analysis, because of the lack
of separation of time scales, and stringent tests will almost be
impossible. In order to overcome these problems we have recently
investigated the relaxation dynamics of a Lennard-Jones system in which
the particles move according to a stochastic dynamics~\cite{gleim98}.
This dynamics leads to a strong damping of the microscopic dynamics and
hence it becomes finally possible to test the predictions of MCT about
the $\beta$-regime and in this paper we report the outcome of these
tests.

The model we investigate is a 80:20 mixture of Lennard-Jones particles
with mass $m$. In the following we will call the two species of
particles A and B. The interaction between two particles of type
$\alpha$ and $\beta$, with $\alpha, \beta \in \{{\rm A,B}\}$, is given by
$V_{\alpha\beta}(r)=4\epsilon_{\alpha\beta}
[(\sigma_{\alpha\beta}/r)^{12} -(\sigma_{\alpha\beta}/r)^6]$ with
$\epsilon_{\rm AA}=1.0$, $\sigma_{\rm AA}=1.0$, $\epsilon_{\rm
AB}=1.5$, $\sigma_{\rm AB}=0.8$, $\epsilon_{\rm BB}=0.5$, and
$\sigma_{\rm BB}=0.88$, and a cut-off radius of
$2.5\sigma_{\alpha\beta}$. In the following we will always use reduced
units with $\sigma_{\rm AA}$ and $\epsilon_{\rm AA}$ the unit of
length and energy, respectively (setting the Boltzmann constant $k_B$
equal to 1.0). Time is measured in units of $\sqrt{\sigma_{\rm
AA}^2m/48\epsilon_{\rm AA}}$. The volume of the simulation box is kept
constant with a box length of 9.4. The dynamics of the system is given
by the stochastic equations of motion
\begin{equation}
m\ddot{{\bf r}}_j+\nabla_j \sum_l
V_{\alpha_j\beta_l}(|{\bf r}_l-{\bf r}_j|) =-\zeta \dot{{\bf r}}_j + 
{\bf \eta}_j(t)\quad.
\label{eq1}
\end{equation}
Here ${\bf \eta}_j(t)$ is a gaussian distributed white noise force with
zero mean.  Because of the fluctuation dissipation theorem, the
magnitude of ${\bf \eta}_j(t)$ is related to $\zeta$ by $\langle
{\bf \eta}_j(t)\cdot {\bf \eta}_l(t') \rangle = 6k_B T \zeta
\delta(t-t')\delta_{jl}$. We have used a value of $\zeta=10$, which is
so large that the presented results for the dynamics do not
depend on $\zeta$ anymore (apart from a trivial change of the time
scale). Equations~(\ref{eq1}) were solved with a Heun algorithm with a
time step of 0.008. The temperatures investigated were 5.0, 4.0, 3.0,
2.0, 1.0, 0.8, 0.6, 0.55, 0.5, 0.475, 0.466, 0.452, and 0.446. At the
lowest temperature the length of the run was $4\times 10^7$ time
steps.  This length is not sufficiently long to equilibrate the
sample.  Therefore we equilibrated the system by means of a Newtonian
dynamics for which we have found that the equilibration times are
significantly shorter~\cite{gleim98}. Thus all the correlation
functions shown in the present work are equilibrium curves, even if
they do not decay to zero at long times. In order to improve the
statistics of the results we averaged at each temperature over 
eight independent runs.

In the following we will review some of the predictions of MCT about
the dynamics in the $\beta$-regime and will denote by $\phi_l(t)$ an
arbitrary time correlation function which couples to density
fluctuations. (Here the index $l$ is just used to distinguish between
different correlators.) As stated here, the predictions are valid only
for temperatures slightly {\it above} the critical temperature $T_c$ of
MCT.  More general results can be found in Refs.~\cite{mct,beta_theo}.

MCT predicts that in the $\beta$-region any correlation function 
$\phi_l(t)$ can be written as
\begin{equation}
\phi_l(t)=f_l^c+h_lc_{\sigma}g_-(t/t_{\sigma})\quad,
\label{eq2}
\end{equation}
where the temperature independent constants $f_l^c$ and $h_l$ are
called critical nonergodicity parameter and critical amplitude, respectively.
The quantity $c_{\sigma}$ is given by 
\begin{equation}
c_{\sigma}=\sqrt{|\sigma|} \qquad \mbox{with} \quad \sigma=C(T_c-T)
\quad,
\label{eq2b}
\end{equation}
where $C$ is a constant. The function $g_-$ is independent of
$\phi_l$ and depends only on the so-called ``exponent parameter''
$\lambda$ which can be calculated from the 
structure factor~\cite{mct}. This calculation has been done for
the present system and a value of $\lambda=0.708$ was
found~\cite{nauroth97}. Hence in our case $\lambda$ is not a fit
parameter. Once $\lambda$ is known, the function $g_-$ can be
calculated numerically.

The quantity $t_{\sigma}$ in Eq.~(\ref{eq2}) is the time scale of the
$\beta$-relaxation and is given by
\begin{equation}
t_{\sigma}=t_0/|\sigma|^{1/2a} \quad,
\label{eq3}
\end{equation}
where $t_0$ is a system universal constant, and the exponent $a$ can be
calculated from $\lambda$ and is in our case
$a=0.324$~\cite{nauroth97}.  Hence, according to MCT, in
Eq.~(\ref{eq2}) only the time scale $t_{\sigma}$ and the prefactor
$h_lc_{\sigma}$ depend on temperature.

MCT also predicts that $\tau_l(T)$, the time scale of the
$\alpha$-relaxation, depends on temperature like
\begin{equation}
\tau_l=\Gamma_l \tau, \quad \tau=t_0/|\sigma|^{\gamma}, \qquad 
\mbox{with}\quad \gamma=1/2a+1/2b
\label{eq4}
\end{equation}
where $\Gamma_l$ is independent of temperature and the exponent $b$ can
also be calculated from $\lambda$~\cite{mct} and is for our system
$b=0.627$~\cite{nauroth97}. Thus we have $\gamma=2.34$.

\begin{figure}[h]
\psfig{file=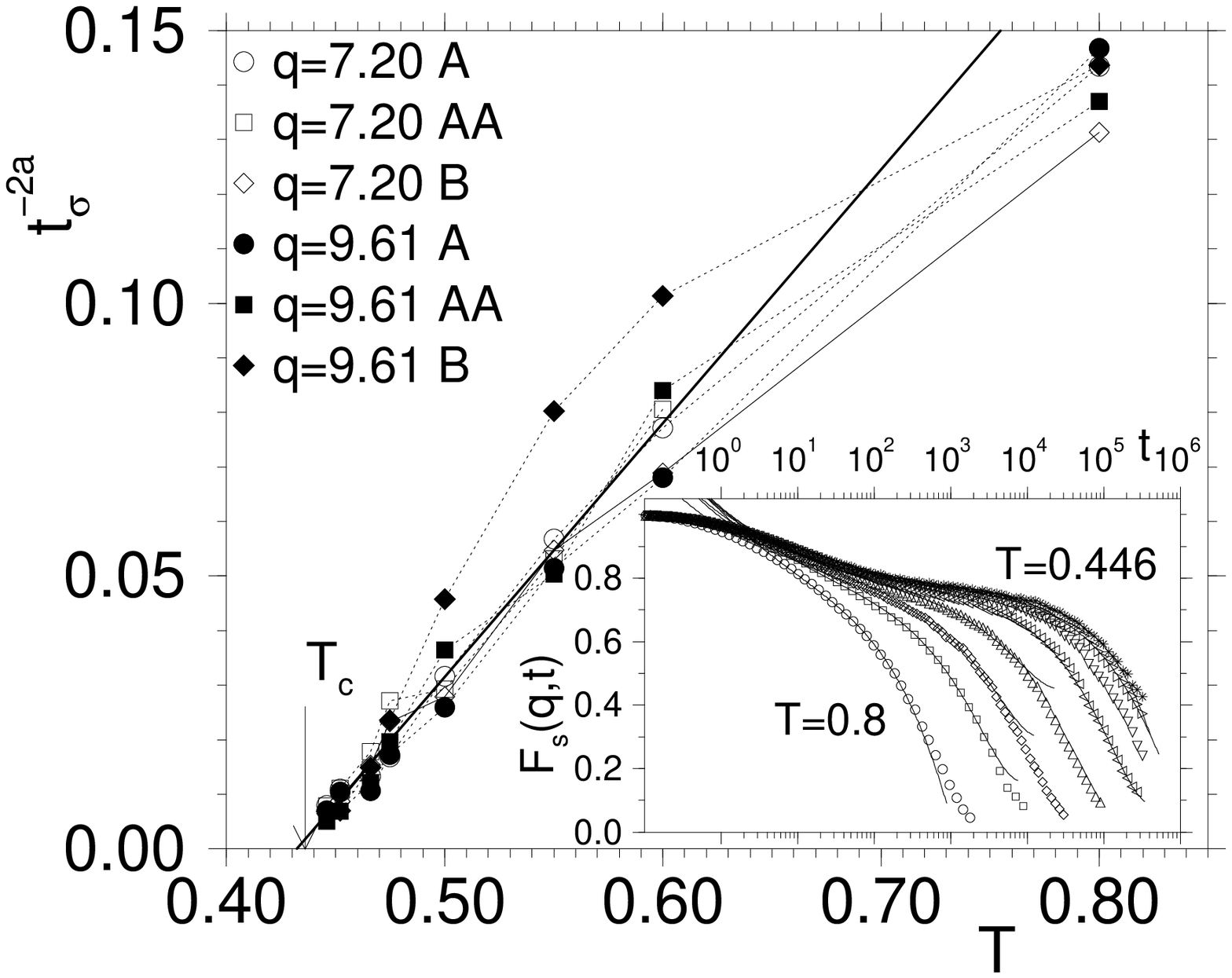,width=11cm,height=8.0cm}
\caption{Main figure: Check of the validity of Eq.~(\protect\ref{eq3})
for various correlators (see labels of curves). MCT predicts a straight
line which intercept the $T-$axis at $T_c=0.435$. The bold straight
line is a linear fit to the open circles. Inset: Time dependence of
$F_s(q,t)$ for the A particles at $q=7.2$ (symbols) and the fitted 
$\beta$-correlators (solid curves) for all $T\leq 0.8$.}
\label{fig1}
\end{figure}

Having presented some of the predictions of MCT about the
$\beta$-relaxation we can now check how well they agree with reality.
For this we calculated from the simulation $F^{\alpha\beta}(q,t)$ and
$F_s^{\alpha}(q,t)$, the coherent and incoherent scattering functions
for wave-vector $q$, respectively. These time correlation functions
were then fitted in the $\beta$-relaxation regime with the functional
form given by Eq.~(\ref{eq2}), where $f_l^c$, $h_lc_{\sigma}$, and
$t_{\sigma}$ were fit parameters. This fit was first done for the
lowest temperature ($T=0.446$). For the fits at the higher temperatures
the value of $f_l^c$ was kept fixed to the one of $T=0.446$ in order to
avoid that the fits give some {\it effective} time scales $t_{\sigma}$
and prefactors $h_lc_{\sigma}$. In the inset of Fig.~\ref{fig1} we
show the results of such fits and it can be seen that the range over
which the $\beta$-correlator describes the data increases with
decreasing temperature, as predicted by MCT.

From Eqs.~(\ref{eq2b}) and (\ref{eq3}) it follows that according to MCT
a plot of $t_{\sigma}^{-2a}$ versus temperature should give a straight
line and that this line should be independent of the correlator
$\phi_l$. In Fig.~\ref{fig1} we show such a plot where we have used for
$\phi_l$ the functions $F_s^{\alpha}(q,t)$ for the A and B particles
and the function $F^{\rm AA}(q,t)$, for two wave-vectors: $q=7.20$ and
$q=9.61$, which correspond to the location of the maximum and first
minimum in the structure factor for the A-A correlation~\cite{kob_lj}.
From this plot we see that at low temperatures the different curves
are indeed close to straight lines and collapse quite well onto a
master curve. Hence we conclude that these two predictions of MCT
work well for our system. Also included in the figure is a linear
fit to the data for $F_s^{\rm A}(q,t)$ for $q=7.2$. This fit
intercepts the temperature axis at $T\approx 0.432$, which according
to MCT should be $T_c$. This estimate of the critical temperature is
in excellent agreement with the one of Ref.~\cite{kob_lj}, where
$T_c=0.435$ was found. 

We also mention that we have found that the square of the prefactor in
Eq.~(\ref{eq2}), $h_lc_{\sigma}$, shows a linear dependence on $T$,
$(h_lc_{\sigma})^2 = |\sigma|$, and vanishes at $T_c$, which follows
from Eqs.~(\ref{eq2}) and (\ref{eq2b}) and is hence in agreement with
MCT. The test of Eq.~(\ref{eq2b}) is equivalent to the test of the relation
between $h_lc_{\sigma}$ and $t_{\sigma}$, which according to
the theory, Eqs.~(\ref{eq2}),(\ref{eq2b}) and (\ref{eq3}), should be

\begin{equation}
h_lc_{\sigma} \propto t_{\sigma}^{-a} \quad.
\label{eq6}
\end{equation}

\begin{figure}[h]
\psfig{file=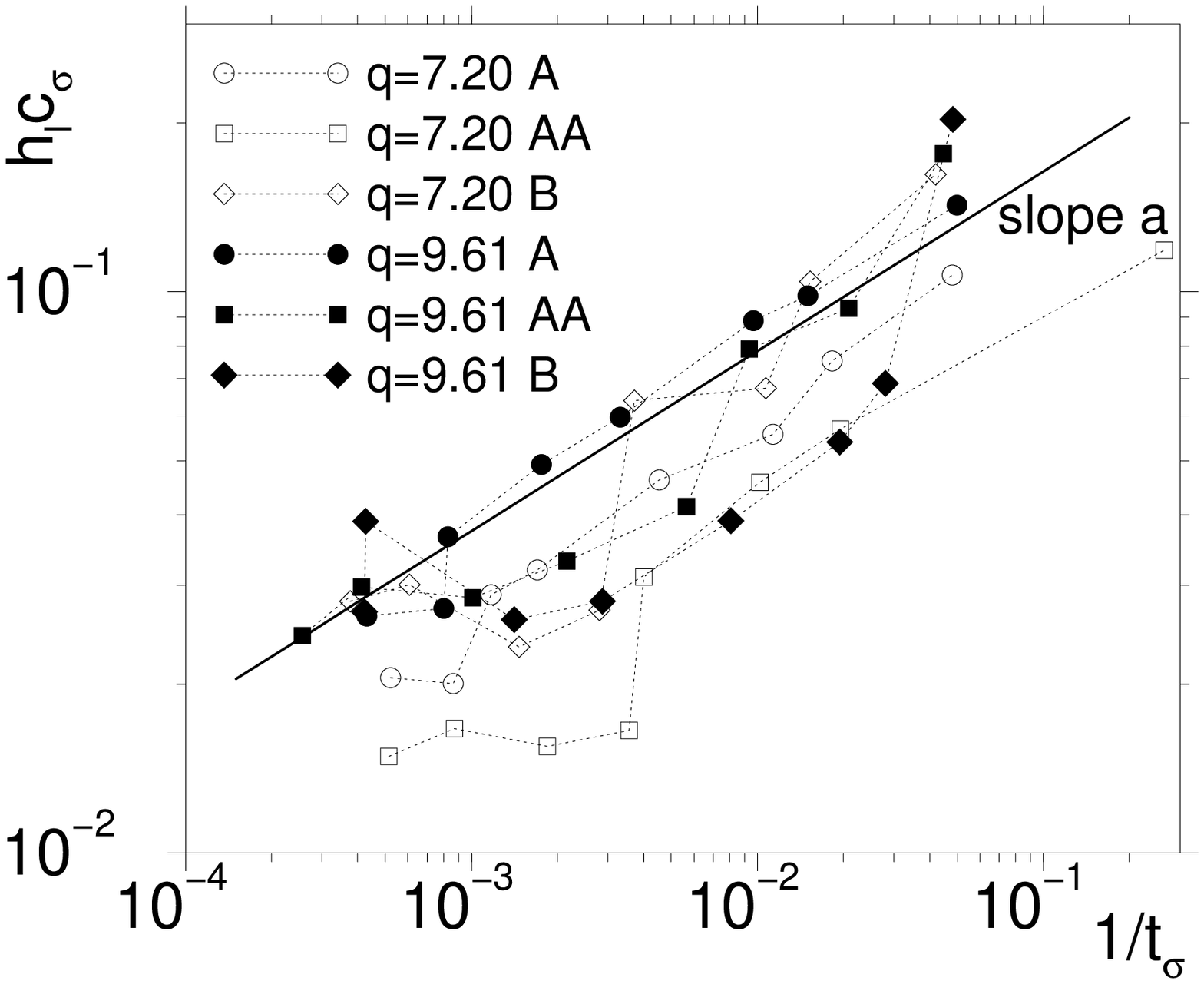,width=11cm,height=8.0cm}
\caption{Check of the validity of Eq.~(\protect\ref{eq6}) for various
correlators. The prediction of MCT is a straight line with slope $a$
(bold straight line).}
\label{fig2}
\end{figure}

In Fig.~\ref{fig2} we plot $h_lc_{\sigma}$ versus $1/t_{\sigma}$ in a
double logarithmic plot for the same correlators discussed in
Fig.~\ref{fig1}. We see that the different curves can be approximated
reasonably well by straight lines with a slope $a$ (bold solid line
in the figure). Therefore we conclude that also this prediction of the
theory seems to work satisfactorily well.

From Eqs.~(\ref{eq3}) and Eqs.~(\ref{eq4}) it follows that also the
$\alpha$-relaxation time $\tau_l$ should show a power-law dependence
on $t_{\sigma}$, i.e. 

\begin{equation}
\tau_l \propto \Gamma_l t_{\sigma}^{1+a/b}\quad . 
\label{eq6b}
\end{equation}

\begin{figure}[h]
\psfig{file=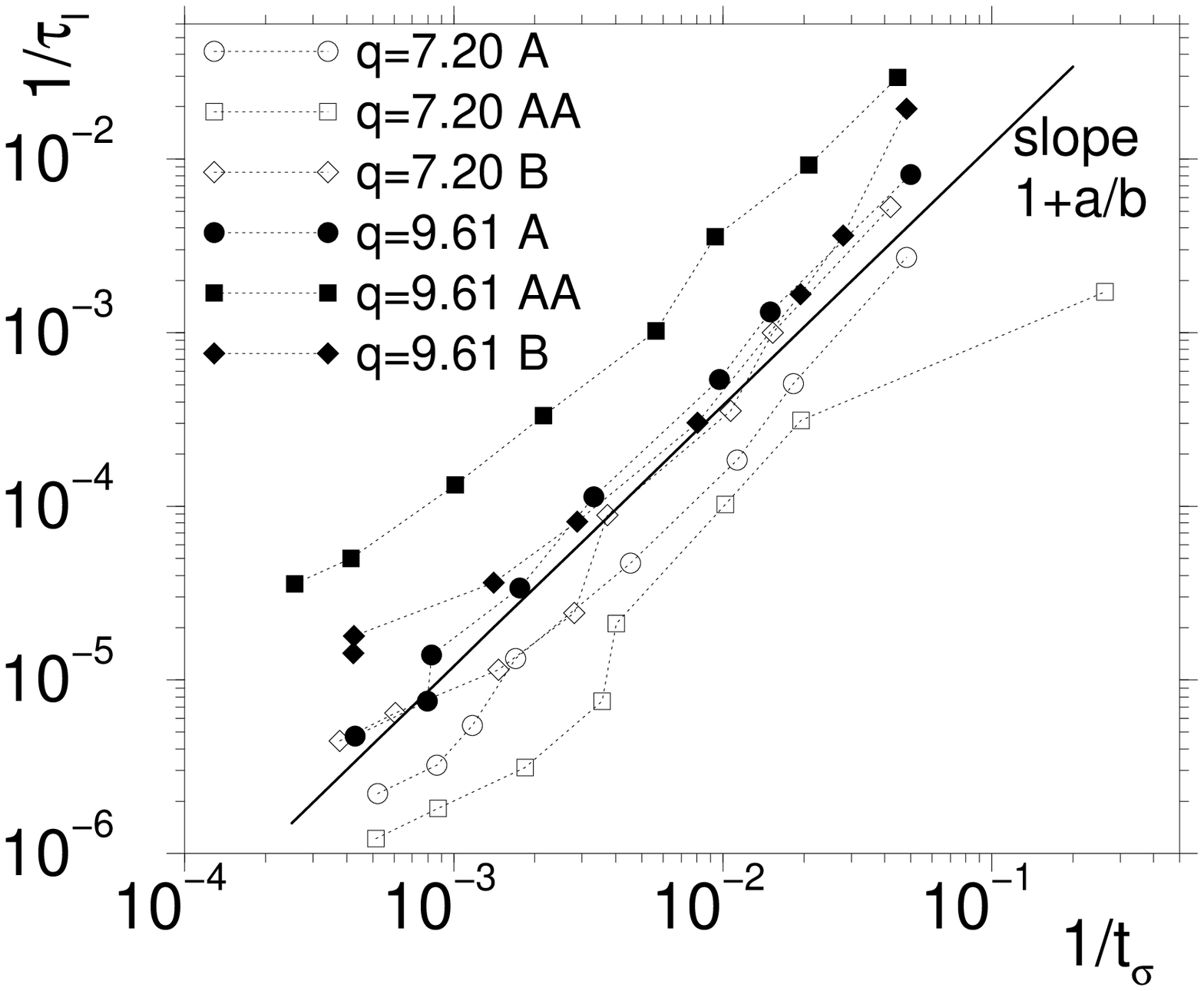,width=11cm,height=8.0cm}
\caption{Check of the validity of Eq.~(\protect\ref{eq6b}) for various
correlators. The prediction of MCT is a straight line with slope $1+a/b$
(bold straight line).}
\label{fig3}
\end{figure}

Thus this equation expresses the surprising prediction of MCT that {\it
two} diverging time scale exist in supercooled liquids, namely $\tau_l$
and $t_{\sigma}$.  Whether this is indeed the case is tested in
Fig.~\ref{fig3} where we plot $\tau_l^{-1}$ versus $t_{\sigma}^{-1}$
for the usual correlators in a double logarithmic plot. We see that the
different curves are indeed close to straight lines and that the slope
is very close to the theoretical value, bold straight line. Thus we
confirm also this prediction of the theory.

\begin{figure}[h]
\psfig{file=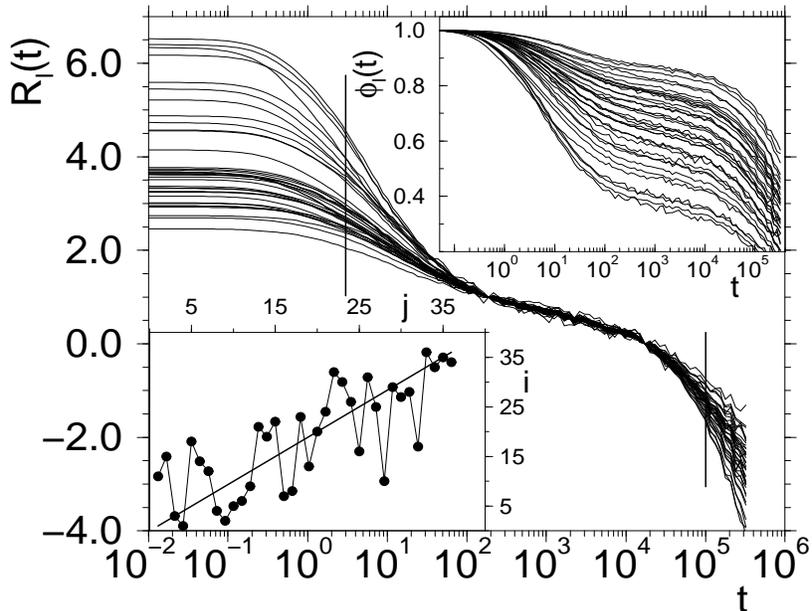,width=11cm,height=8.0cm}
\caption{Main figure: Time dependence of the ratio given in
Eq.~(\protect\ref{eq7}), demonstrating the validity of the
factorization property [ Eq.~(\protect\ref{eq2})]. $T=0.446$. The
correlation functions $\phi_l(t)$ are shown in the upper right inset.
See text for the discussion of the lower left inset and the two
vertical lines in the main figure.}
\label{fig4}
\end{figure}

As already mentioned above in the context of Eq.~(\ref{eq2}), according
to MCT the whole time dependence of $\phi_l(t)$ is given by the
$l-$independent function $g_-(t/t_{\sigma})$. In order to test this
prediction we can introduce a function $R_l(t)$ as follows:
\begin{equation}
R_l(t)=\frac{\phi_l(t)-\phi_l(t')}{\phi_l(t'')-\phi_l(t')}\quad .
\label{eq7}
\end{equation}
Here $t'$ and $t''$ are arbitrary times in the $\beta$-relaxation
regime ($t'\neq t''$). From Eq.~(\ref{eq2}) it follows immediately that
in the $\beta$-regime the function $R_l(t)$ is independent of the
correlator, i.e. of $l$. To see whether this is indeed the case we have
considered the correlation function discussed in the context of
Fig.~\ref{fig1} and in addition the coherent and incoherent scattering
function for several other wave vectors and also the cross correlation
function $F^{\rm AB}(q,t)$ at different $q$. This gave us a total of 36
correlation functions which are shown in the upper inset of
Fig.~\ref{fig4} (at $T=0.446$). For each of these functions we
determined the corresponding $R_l(t)$, choosing for $t'$ and $t''$ a
value around 200 and 15000, respectively. In the main figure of
Fig.~\ref{fig4} we show the different $R_l(t)$ and we see that in the
$\beta$-regime they do indeed collapse onto a master curve. That such a
collapse is not a trivial result can be concluded from the observation
that outside the $\beta$-regime the different curves show a strong
dependence on $l$, at short as well as at long times.

Equation (\ref{eq2}) is the prediction of the theory about the {\it
leading} asymptotic behavior for the time and temperature dependence of
a generic correlator. Very recently the next order corrections to this
behavior have been calculated~\cite{mct_corrections} and these 
corrections can now be used to
do more checks on the validity of the theory. In
Ref.~\cite{mct_corrections} it has been shown that in the {\it early}
$\beta$-relaxation regime, i.e. for $t_0 \ll t \ll t_{\sigma}$, the
correlator can be written as
\begin{equation}
\phi_l(t)=f_l^c + h_l (t_0/t)^a \{ 1 + [K_l+\Delta] (t_0/t)^a\} \quad .
\label{eq8}
\end{equation}
Here $\Delta$ is a $l-$independent constant and the constant $K_l$
depends on $l$ but not on temperature. In the {\it late}
$\beta$-regime, for which $t_{\sigma}\ll t\ll \tau_l$, the correlation
function is predicted to behave like

\begin{equation}
\phi_l(t)=f_l^c - h_l (t/\tau)^b \{ 1 -K_l (t/\tau)^b \} \quad .
\label{eq9}
\end{equation}

The mentioned corrections are the second terms in the curly brackets
in Eqs.~(\ref{eq8}) and (\ref{eq9}).
The important result about these equations is that the
$l$ dependent part of the correction, i.e. $K_l$, is the same. Using
this fact it is simple to show the following: Calculate the ratio
$R_l(t)$ from Eq.~(\ref{eq7}) for various correlators and plot these
$R_l(t)$ versus the logarithm of $t$. Draw two vertical lines at times
that are a bit shorter and a bit longer than the times where the
asymptotic expression, Eq.~(\ref{eq2}), holds. Start to label the
correlators from top to bottom in the order they intersect the vertical
line at {\it short} times and call this number $i$. Determine the position
$j$ at which curve $i$ intersects the vertical line at {\it large} times,
{\it where the counting is again done from top to bottom}. Thus this
gives a function $j(i)$. From Eqs.~(\ref{eq8}) and (\ref{eq9}) it then
follows that $j=i$. Or to put this in other words: the first (second,
\ldots) curve that intersects the left vertical line is also the first
(second, \ldots) curve to intersect the right vertical line.

We have done the described procedure by using vertical lines at $t=3$
and $t=10^5$ (bold vertical lines in Fig.~\ref{fig4}). The function
$j(i)$ we find is shown in the lower inset of Fig.~\ref{fig4}. We see
that, despite the scattering present in the data, a clear increasing
trend which is compatible with a straight line with unit slope can be
seen, thus giving also support for the validity of this prediction of
the theory.

We thus can conclude from the present work that many of the predictions
that mode-coupling theory makes for the $\beta$-relaxation can also be
tested in computer simulations. As we have shown in this paper the
outcome of such tests for the Lennard-Jones system considered here is
that the theory is indeed able to give a self consistent picture of the
dynamics of this simple glass-former in the $\beta$-relaxation regime.

Acknowledgements: We thank W. G\"otze and A. Latz for valuable
discussions and comments on a draft of this paper. Part of this work
was supported by the DFG under SFB 262/D1.

\end{document}